\documentclass[aps,prb,twocolumn,showpacs,byrevtex]{revtex4}

\usepackage{times,mathptm,gensymb}
\usepackage[dvips]{graphicx,color}

\begin{document}

\title{Equivalence of Electronic and Mechanical Stresses in Structural Phase Stabilization: A Case Study of Indium Wires on Si(111)}
\author{Sun-Woo Kim$^1$, Hyun-Jung Kim$^1$, Fangfei Ming$^{3,4}$, Yu Jia$^2$, Changgan Zeng$^{3,4,5*}$, Jun-Hyung Cho$^{1,2{\dagger}}$, and Zhenyu Zhang$^{4,5}$}
\affiliation{$^1$ Department of Physics and Research Institute for Natural Sciences, Hanyang University, 17 Haengdang-Dong, Seongdong-Ku, Seoul 133-791, Korea \\
$^2$ Center for Clear Energy and Quantum Structures, and School of Physics and Engineering, Zhengzhou University, Zhengzhou 450052, China \\
$^3$ Hefei National Laboratory for Physical Sciences at the Microscale (HFNL), CAS Key Laboratory of Strongly-Coupled Quantum Matter Physics, and Department of Physics, University of Science and Technology of China, Hefei, Anhui 230026, China \\
$^4$ International Center for Quantum Design of Functional Materials (ICQD), HFNL,
University of Science and Technology of China, Hefei, Anhui 230026, China \\
$^5$ Synergetic Innovation Center of Quantum Information and Quantum Physics,
University of Science and Technology of China, Hefei, Anhui 230026, China
}

\date{\today}

\begin{abstract}
It was recently proposed that the stress state of a material can also be altered via electron or hole doping, a concept termed electronic stress (ES), which is different from the traditional mechanical stress (MS) due to lattice contraction or expansion. Here we demonstrate the equivalence of ES and MS in structural stabilization, using In wires on Si(111) as a prototypical example. Our systematic density-functional theory calculations reveal that, first, for the same degrees of carrier doping into the In wires, the ES of the high-temperature metallic 4${\times}$1 structure is only slightly compressive, while that of the low-temperature insulating 8${\times}$2 structure is much larger and highly anisotropic. As a consequence, the intrinsic energy difference between the two phases is significantly reduced towards electronically phase-separated ground states. Our calculations further demonstrate quantitatively that such intriguing phase tunabilities can be achieved equivalently via lattice-contraction induced MS in the absence of charge doping. We also validate the equivalence through our detailed scanning tunneling microscopy experiments. The present findings have important implications in understanding the underlying driving forces involved in various phase transitions of simple and complex systems alike.
\end{abstract}


\maketitle

\vspace{0.4cm}
\section{INTRODUCTION}
\vspace{0.4cm}

Mechanical stress (MS) produced by lattice deformation is well established to tune the electronic, magnetic, optical, and phononic properties of materials~\cite{Camm}, and such an elastic stress (strain) engineering has been widely adopted to substantially improve the carrier mobilities in semiconductor devices~\cite{Schaffler}. These MS-driven tuning effects are more profound in strongly-correlated or low-dimensional materials, mainly due to the enhanced entanglement between lattice, charge, spin, and orbit degrees of freedom. For example, MS has been demonstrated to tune the Mott transition temperature in VO$_2$ nanowires~\cite{JISohn}, and to generate giant pseudomagnetic field and band gap in graphene~\cite{Guinea,Pereira}. The creation of MS usually relies on high-pressure instruments or lattice mismatch engineering at the interfaces. Contrasting with the MS induced by lattice deformation, the so-called quantum electronic stress (ES), a pure electronic effect on the stress originating from the variation of carrier density, has been recently introduced and formulated within density functional theory (DFT)~\cite{Hu}. Indeed, the ES induced by quantum electronic confinement in metal thin films has been demonstrated theoretically~\cite{Hu,Huang,MLiu} and experimentally~\cite{Flototto}. Since the MS and ES have substantially different origins involving explicitly the variations of lattice and charge degrees of freedom, respectively, it is interesting and challenging to examine whether and how they can equivalently tune the physical properties, especially in the same system.

For the surface structures formed by epitaxial metal atom adsorption on semiconductor surfaces, there are frequently competing electronic phases~\cite{Carpinelli,Tejeda,Polei,Zhang3} because of their reduced phase space, and the stability of these phases can be effectively tuned by deforming the lattice~\cite{Zhang3} or by varying charge carriers~\cite{Polei}. In this sense, the low-dimensional electronic systems formed on surfaces provide a unique playground to demonstrate the tuning effect of phase stability in terms of surface MS and ES (hereafter MS and ES refer to the surface ones). Here, we focus on a prototypical example of quasi-one dimensional (1D) systems, self-assembled Indium (In) atom wires on the Si(111) surface (see Fig. 1). This In/Si(111) surface system undergoes a structural phase transition from a high-temperature metallic 4${\times}$1 phase [see Fig. 1(a)] to a low-temperature insulating 8${\times}$2 phase [Fig. 1(b)] at a transition temperature ($T_{\rm c}$) of $-$125 K~\cite{Yeom}. The structural model of the 8${\times}$2 phase is well established to have the formation of In hexagons via a periodic lattice distortion~\cite{gon1,gon2,HJKim}, reflecting the presence of the MS. To produce and quantify the ES that may influence the relative stability of the 4${\times}$1 and 8${\times}$2 structures, we introduce electron doping into In wires. It is very interesting to explore how such an electron-doping induced ES changes depending on the metallic and insulating phases, and also to examine the equivalence of the ES and MS in tuning the phase stabilization of the 4${\times}$1 and 8${\times}$2 structures.

\begin{figure}[ht]
\centering{ \includegraphics[width=7.7cm]{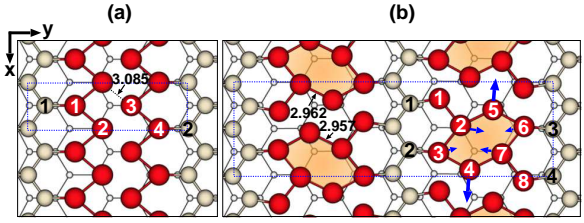} }
\caption{(Color on line) Top view of the optimized (a) 4${\times}$1 and (b) 8${\times}$2 structures of In/Si(111). Each In atom wire is composed of two zigzag chains of In atoms. The dark and gray circles represent In and Si atoms, respectively. For distinction, Si atoms in the subsurface are drawn with small circles. Each unit cell is indicated by the dotted line. The ${\bf x}$ (${\bf y}$) axis is parallel (perpendicular) to In chains. Numbers denote the In-In distance (in {\AA}) between In chains. The arrows in (b) indicate schematically the relaxation directions of In atoms forming the In hexagon (see Table I), when electrons are doped to In wires.}
\end{figure}

In this paper, we present a comprehensive study of surface MS and ES in the In/Si(111) system by using a van der Waals (vdW) energy-corrected hybrid DFT calculation. We find that the formation of In hexagons brings a significant reduction of the tensile MS perpendicular to In wires, leading to the stabilization of the 8${\times}$2 structure. Interestingly, the ES induced by electron doping into In wires exhibits drastically different features between the 4${\times}$1 and 8${\times}$2 structures: i.e., the ES of the 4${\times}$1 metallic structure is slightly compressive, while that of the 8${\times}$2 insulating structure is anisotropic with a highly compressive (tensile) component along the direction parallel (perpendicular) to In wires. As a result, the surface energy difference between the 4${\times}$1 and 8${\times}$2 structures decreases as the amount of electron doping increases. This ES-driven tuning effect on the relative stability of the 4${\times}$1 and 8${\times}$2 structures is found to be equivalent to the MS-driven one obtained by applying a compressive lattice strain. Our theoretical predictions are confirmed by scanning tunneling microscopy (STM) measurement at 5 K.

\vspace{0.4cm}
\section{CALCULATIONAL METHOD}
\vspace{0.4cm}

The present hybrid DFT+vdW calculation~\cite{TS-vdW,Zhang-vdW} was performed using the FHI-aims~\cite{Aims} code for an accurate, all-electron description based on numeric atom-centered orbitals, with ``tight" computational settings. For the exchange-correlation energy, we employed the screened hybrid functional of Heyd-Scuseria-Ernzerhof (HSE)~\cite{HSE1,HSE2}. The Si(111) substrate (with the Si lattice constant $a_0$ = 5.418 {\AA}) was modeled by a 6-layer slab (not including the Si surface chain bonded to the In chains) with ${\sim}$30 {\AA} of vacuum in between the slabs, where each Si atom in the bottom layer was passivated by one H atom. The ${\bf k}$-space integrations were done equivalently with 64 and 16 ${\bf k}$ points in the surface Brillouin zone of the 4${\times}$1 and 8${\times}$2 unit cells, respectively. All atoms except the bottom layer were allowed to relax along the calculated forces until all the residual force components were less than 0.001 eV/{\AA}. The employed hybrid DFT+vdW scheme was successfully applied not only to determine the energy stability of the 4${\times}$1 and 8${\times}$2 structures of In/Si(111)~\cite{HJKim} but also to calculate the stress tensor~\cite{Aims-stress}.

\vspace{0.4cm}
\section{RESULTS}
\vspace{0.4cm}

We begin to optimize the 4${\times}$1 and 8${\times}$2 structures without electron doping by using the hybrid DFT+vdW scheme. The optimized 4${\times}$1 and 8${\times}$2 structures are displayed in Fig. 1(a) and 1(b), respectively. It is seen that the 8${\times}$2 structure has the shorter In-In distances ($d_{\rm In-In}$ = 2.957 and 2.962 {\AA}) between two In chains compared to that (3.085 {\AA}) in the 4${\times}$1 structure, forming In hexagons. Such an 8${\times}$2 hexagon structure is found to be more stable than the 4${\times}$1 structure by 33 meV per 4${\times}$1 unit cell. The calculated surface band structures of the 4${\times}$1 and 8${\times}$2 structures show that the 4${\times}$1 structure exhibits the presence of three metallic bands crossing the Fermi level whereas the 8${\times}$2 structure has a band gap of 0.31 eV, in good agreement with previous experimental data~\cite{Tanikawa,SJPark,Zhang2}. To examine how the MS changes after the formation of In hexagons, we calculate the MS difference ${\Delta}{\sigma}_{ij}^{\rm M}$ between the 8${\times}$2 (${\alpha}$) and 4${\times}$1 (${\beta}$) structures, defined as
\begin{eqnarray}
{\Delta}{\sigma}_{ij}^{\rm M} = {\sigma}_{ij,{\alpha}}^{\rm M} - {\sigma}_{ij,{\beta}}^{\rm M} = \frac{1}{A_{\alpha}} \frac{{\partial}(A_{\alpha}{\gamma}_{\alpha})}{{\partial}{\epsilon}_{ij}} - \frac{1}{A_{\beta}} \frac{{\partial}(A_{\beta}{\gamma}_{\beta})}{{\partial}{\epsilon}_{ij}} \nonumber\\
= \frac{1}{A_{\alpha}} \frac{{\partial}E_{slab,\alpha}}{{\partial}{\epsilon}_{ij}} - \frac{1}{A_{\beta}} \frac{{\partial}E_{slab,\beta}}{{\partial}{\epsilon}_{ij}}.
\end{eqnarray}
Here, ${\epsilon}_{ij}$ ($i,j$ = $x,y$) denotes the element of strain tensor, $A$ the surface area, and ${\gamma}$ ($E_{slab}$) the surface (slab) energy. For the last equality, see the Appendix. Therefore, ${\Delta}{\sigma}_{ij}^{\rm M}$ can be evaluated by using $E_{slab}$ obtained from the slab calculation. The calculated results of ${\Delta}{\sigma}_{ij}^{\rm M}$ are plotted in Fig. 2(a), together with those (discussed below) obtained with electron doping. We find that (i) the ${\sigma}_{xx}^{\rm M}$ and ${\sigma}_{xy}^{\rm M}$ components in 8${\times}$2 are almost the same as those in 4${\times}$1 and (ii) the ${\sigma}_{yy}^{\rm M}$ component in 8${\times}$2 is reduced as much as 29.79 meV/{\AA}$^2$ compared to that in 4${\times}$1. Thus, the In-hexagon formation results in a decrease in the tensile surface stress along the $y$ direction, giving rise to the stabilization of the 8${\times}$2 structure. Here, we note that the absolute value of ${\sigma}_{ij}^{\rm M}$ in the 4${\times}$1 reference is ${\sigma}_{xx}^{\rm M}$ = 54.79, ${\sigma}_{yy}^{\rm M}$ = 121.68, and ${\sigma}_{xy}^{\rm M}$ = 0 meV/{\AA}$^2$, indicating a tensile surface stress along the $x$ and $y$ directions.

\begin{figure*}[ht]
\centering{ \includegraphics[width=17.4cm]{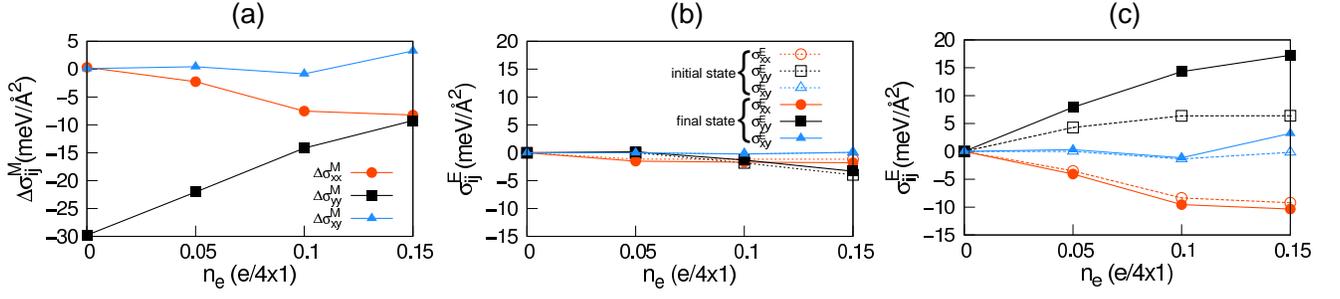} }
\caption{(Color on line) (a) Calculated MS difference ${\Delta}{\sigma}_{ij}^{\rm M}$ between the 8${\times}$2 and 4${\times}$1 structures as a function of $n_e$. The ES ${\sigma}_{ij}^{\rm E}$ for the 4${\times}$1 and 8${\times}$2 structures are given in (b) and (c), respectively.}
\end{figure*}

Next, we study the ES induced by electron doping with excess electronic charge $n_e$ per 4${\times}$1 unit cell~\cite{VCM}. The concept of ES was recently formulated within DFT~\cite{Hu}, and it can be practically calculated by using the difference of the MS obtained at the total electronic charge $n_d$ = $n_0$ + $n_e$ and that at the ground-state electronic charge $n_0$:
\begin{eqnarray}
{\sigma}_{ij}^{\rm E}(n_e)  = {\sigma}_{ij}^{\rm M}(n_d) - {\sigma}_{ij}^{\rm M}(n_0) = \frac{1}{A} \frac{{\partial}E_{slab}(n_d)}{{\partial}{\epsilon}_{ij}} - \frac{1}{A} \frac{{\partial}E_{slab}(n_0)}{{\partial}{\epsilon}_{ij}}.
\end{eqnarray}
We here consider the two different states for the treatment of ${\sigma}_{ij}^{\rm M}(n_d)$: one is the ``initial" state without the relaxation of atoms (i.e., fixing the structure having $n_0$) and the other is the ``final" state which allows the atomic relaxation along the generated forces due to electron doping. The calculated initial-state and final-state ES results for the 4${\times}$1 and 8${\times}$2 structures are plotted as a function of $n_e$ in Fig. 2(b) and 2(c), respectively. For the 4${\times}$1 initial state, ${\sigma}_{xx}^{\rm E}$ (${\sigma}_{yy}^{\rm E}$) is slightly negative as $-$1.15 ($-$0.01), $-$1.15 ($-$1.81), and $-$1.16 ($-$3.96) meV/{\AA}$^2$ for $n_e$ = 0.05, 0.1, and 0.15$e$, respectively. The inclusion of lattice relaxation within the 4${\times}$1 final state shows a negligible change in ${\sigma}_{ij}^{\rm E}$ [see Fig. 2(b)]. Thus, we can say that electron doping in the 4${\times}$1 structure produces a weakly compressive ES. Contrasting with the 4${\times}$1 case, the 8${\times}$2 initial state exhibits larger electron-doping effects with ${\sigma}_{xx}^{\rm E}$ (${\sigma}_{yy}^{\rm E}$) = $-$3.50 (+4.29), $-$8.34 (+6.38), and $-$9.22 (+6.39) meV/{\AA}$^2$ for $n_e$ = 0.05, 0.1, and 0.15$e$, respectively. As shown in Fig. 2(c), the 8${\times}$2 final state further increases the magnitude of ${\sigma}_{xx}^{\rm E}$ (${\sigma}_{yy}^{\rm E}$) as $-$4.07 (+7.91), $-$9.55 (+14.30), and $-$10.34 (+17.25) meV/{\AA}$^2$ for $n_e$ = 0.05, 0.1, and 0.15$e$, respectively, and their magnitudes monotonically increase with increasing $n_e$. Noting that the In-hexagon formation in the 8${\times}$2 structure involves a reduction of the mechanical tensile surface stress [see Fig. 2(a)], the significant final-state effect of ${\sigma}_{yy}^{\rm E}$ may accompany a large atomic relaxation. Indeed, Table I shows that the electron doping of $n_e$ = 0.1$e$ in the 8${\times}$2 structure gives a conspicuous relaxation of In atoms forming the In hexagon, as indicated by the arrows in Fig. 1(b). On the other hand, for the 4${\times}$1 structure, 
\begin{table}[ht]
\caption{Calculated displacements (in \AA) of In and Si atoms in the 4${\times}$1 and 8${\times}$2 final-state structures with $n_e$ = 0.1$e$ relative to the positions obtained from the corresponding structures without electron doping. The labeling of In and Si atoms is shown in Fig. 1.}
\begin{ruledtabular}
\begin{tabular}{lrrrrrr}
 & & 4${\times}$1 & & & 8${\times}$2 &  \\
 & ${\Delta}x$   & ${\Delta}y$  & ${\Delta}z$  & ${\Delta}x$   & ${\Delta}y$  & ${\Delta}z$  \\ \hline
In$_1$     &  0.000 &    0.001 & $-$0.007 &    0.043 &    0.000 & $-$0.010 \\
In$_2$     &  0.000 & $-$0.002 &  0.007   &    0.009 &    0.071 &    0.022 \\
In$_3$     &  0.000 & $-$0.006 &  0.009   & $-$0.011 &    0.046 & $-$0.025 \\
In$_4$     &  0.000 & $-$0.008 & $-$0.014 &    0.132 & $-$0.011 &    0.003 \\
In$_5$     &        &          &          & $-$0.124 &    0.017 &    0.008 \\
In$_6$     &        &          &          &    0.010 & $-$0.044 & $-$0.032 \\
In$_7$     &        &          &          & $-$0.008 & $-$0.062 &    0.024 \\
In$_8$     &        &          &          & $-$0.044 &    0.006 & $-$0.006 \\
Si$_1$     &  0.000 & $-$0.004 & $-$0.001 &    0.010 & $-$0.005 & $-$0.002 \\
Si$_2$     &  0.000 & $-$0.003 & $-$0.004 & $-$0.006 &    0.015 &    0.010 \\
Si$_3$     &        &          &          & $-$0.008 & $-$0.015 &    0.006 \\
Si$_4$     &        &          &          &    0.003 &    0.014 &    0.001 \\
\end{tabular}
\end{ruledtabular}
\end{table}
there is a negligible atomic relaxation caused by electron doping (see Table I). It is remarkable that the ES of the 8${\times}$2 structure is anisotropic with a highly compressive (tensile) component along the direction parallel (perpendicular) to In wires, and thus their magnitudes are significantly larger than those of the 4${\times}$1 structure.

To account for the different features of ES between the 4${\times}$1 and 8${\times}$2 structures, we display in Fig. 3 the charge characters of their final states with $n_e$ = 0.1$e$, together with the corresponding band structures. For the 4${\times}$1 structure, electron doping shifts the Fermi level continuously upward by occupying the metallic states, and the occupied excess electrons are found to be well distributed over the whole In wires [see Fig. 3(a)], indicating a widely delocalized metallic character. On the other hand, for the 8${\times}$2 structure, the excess electrons occupying the conduction bands above the band gap show a strongly delocalized character along each chain with charge depletion between In chains [see Fig. 3(b)], leading to enhance the metallic bonding along each chain 
\begin{figure}[ht]
\centering{ \includegraphics[width=7.7cm]{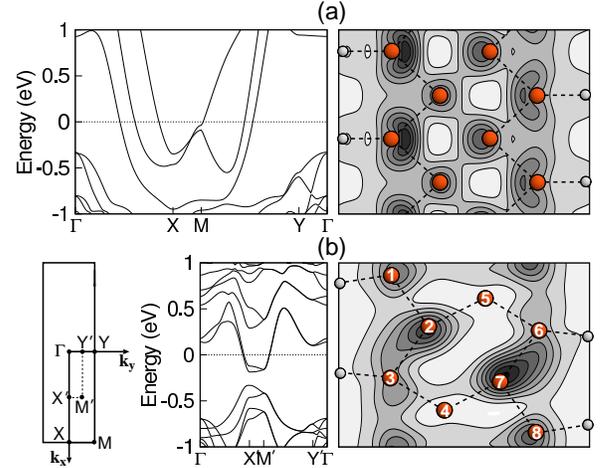} }
\caption{(Color on line) Calculated surface band structures of the electron-doped (a) 4${\times}$1 and (b) 8${\times}$2 final states with $n_e$ = 0.1$e$ per 4${\times}$1 unit cell. The energy zero represents the Fermi level $E_F$. The surface Brilloiun zone is displayed in (b). For each structure, the charge character of excess electrons, obtained by the charge density difference ${{\rho}_{n_d}}-{{\rho}_{n_0}}$, is also given. Here, the charge contour plot with a contour spacing of 0.2${\times}$10$^{-3}$e/{\AA}$^3$ is drawn in a lateral plane near In atom wires.}
\end{figure}
\begin{figure}[ht]
\centering{ \includegraphics[width=7.7cm]{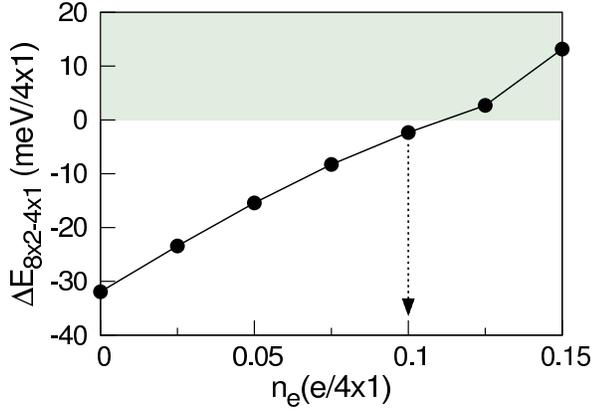} }
\caption{(Color on line) Calculated total-energy difference ${\Delta}E_{\rm 8{\times}2-4{\times}1}$ per 4${\times}$1 unit cell between the 8${\times}$2 and 4${\times}$1 structures as a function of electron doping $n_e$. The arrow indicates the heavy electron doping of ${\sim}$0.1$e$ in the STM and angle-resolved photoemission spectroscopy~\cite{Morikawa}.}
\end{figure}
while to weaken the strength of covalently bound In hexagons formed by In$_2$-In$_5$ and In$_4$-In$_7$ covalent bonds~\cite{HJKim}. Based on such contrasting charge characters of excess electrons between the 4${\times}$1 and 8${\times}$2 structures, it is likely that the 4${\times}$1 structure has a weakly compressive ES, while the 8x2 structure has an anisotropic feature of ES with a highly compressive (tensile) stress along the $x$ ($y$) direction. Here, the electron-doped 8${\times}$2 structure, which weakens the covalent bonding of In atoms between In chains, in turn gives an increase of tensile stress along the $y$ direction.

By summation of ${\sigma}_{ij}^{\rm M}(n_0)$ and ${\sigma}_{ij}^{\rm E}(n_e)$, we can obtain the MS ${\sigma}_{ij}^{\rm M}(n_d)$ at $n_d$ = $n_0$ + $n_e$. Accordingly, the difference of MS between the electron-doped 8${\times}$2 and 4${\times}$1 structures is given by ${\Delta}{\sigma}_{ij}^{\rm M}(n_d)$ = ${\Delta}{\sigma}_{ij}^{\rm M}(n_0)$ + ${\Delta}{\sigma}_{ij}^{\rm E}(n_e)$. The calculated results for ${\Delta}{\sigma}_{ij}^{\rm M}(n_d)$ are plotted as a function of $n_e$ in Fig. 2(a). It is seen that the slope of increase in ${\Delta}{\sigma}_{yy}^{\rm M}$ is greater than that of decrease in ${\Delta}{\sigma}_{xx}^{\rm M}$, thereby giving rise to a decrease in the magnitude of ${\Delta}{\sigma}_{xx}^{\rm M}$+${\Delta}{\sigma}_{yy}^{\rm M}$ with increasing $n_e$. Consequently, one expects a reduction of the total-energy difference ${\Delta}E_{\rm 8{\times}2-4{\times}1}$ (per 4${\times}$1 unit cell) between the 8${\times}$2 and 4${\times}$1 structures with increasing $n_e$. Indeed, as shown in Fig. 4, the present calculation of ${\Delta}E_{\rm 8{\times}2-4{\times}1}$ shows that the relative stability of 8${\times}$2 with respect to 4${\times}$1 decreases with increasing $n_e$. Interestingly, the 4${\times}$1 structure becomes more stabilized than the 8${\times}$2 structure above $n_e$ ${\simeq}$ 0.11$e$, implying that the ground state of the In/Si(111) system can be switched to the 4${\times}$1 structure by electron doping.

\begin{figure}[ht]
\centering{ \includegraphics[width=7.7cm]{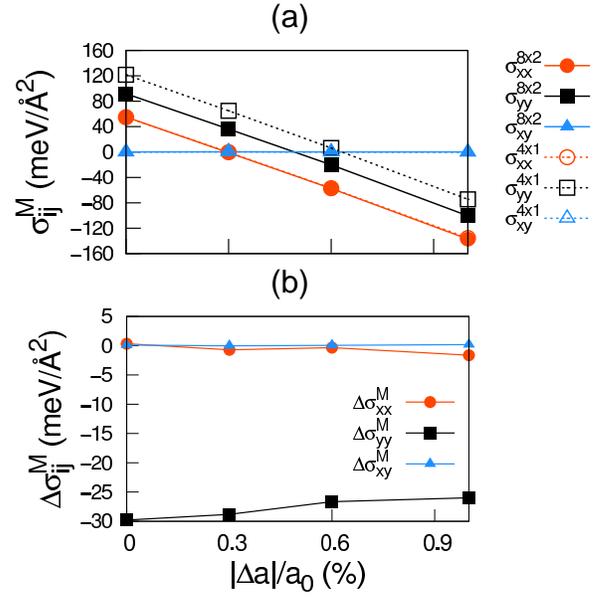} }
\caption{(Color on line) (a) Calculated MS components ${\sigma}_{xx}^{\rm M}$, ${\sigma}_{xy}^{\rm M}$, and ${\sigma}_{yy}^{\rm M}$ for the 4${\times}$1 and 8${\times}$2 structures and (b) MS difference ${\Delta}{\sigma}_{ij}^{\rm M}$ between the 8${\times}$2 and 4${\times}$1 structures as a function of $|{\Delta}a|/a_0$.}
\end{figure}

It is noteworthy that the decrease in the magnitude of ${\Delta}E_{\rm 8{\times}2-4{\times}1}$ with electron doping is consistent with several previous experimental observations that (i) $n$-type Si(111) substrate yields the coexistence of the 4${\times}$1 and 8${\times}$2 phases even at 47 K, whose areal ratio can be tuned by optical excitation that decreases the amount of electron doping in In wires~\cite{Terada} and (ii) electron doping via Na adsorption on the In/Si(111) surface suppresses the (4${\times}$1)${\leftrightarrow}$(8${\times}$2) phase transition, resulting in a lowering of $T_c$~\cite{Morikawa}. Here, electron doping with the Na coverage of ${\sim}$0.02 ML preserved the 4${\times}$1 phase even at 50 K. Remarkably, this Na coverage was estimated to give $n_e$ ${\approx}$ 0.1$e$~\cite{Morikawa}, at which our calculated value of ${\Delta}E_{\rm 8{\times}2-4{\times}1}$ approaches zero (see Fig. 4)~\cite{SCL}.

For comparison with the features of ES, we further study the MS induced by lattice deformation. According to our recent STM and DFT study~\cite{Zhang3}, the vacancy creation in In wires produces a compressive lattice strain to change the relative stability of the 4${\times}$1 and 8${\times}$2 structures, as discussed below. We here examine how the MS evolves with contracting the lattice constant $a$ of the Si(111) substrate by 1\%. As shown in Fig. 5(a), we find that the tensile MS components ${\sigma}^{\rm M}_{xx}$ and ${\sigma}^{\rm M}_{yy}$ in the 4${\times}$1 (8${\times}$2) structures decrease with contracting $a$ and are finally converted to be compressive at a contraction of ${\sim}$0.3(0.3) and ${\sim}$0.6(0.5)\%, respectively. Interestingly, the MS difference ${\Delta}{\sigma}_{xx}^{\rm M}$ (${\Delta}{\sigma}_{yy}^{\rm M}$) between the 8${\times}$2 and 4${\times}$1 structures decreases (increases) with contracting $a$ [see Fig. 5(b)], similar to the pattern of ${\Delta}{\sigma}^{\rm M}_{xx}$ (${\Delta}{\sigma}^{\rm M}_{yy}$) as a function of $n_e$ in the above-mentioned case of electron doping [see Fig. 2(a)]. The resulting magnitude of ${\Delta}{\sigma}^{\rm M}_{xx}$+${\Delta}{\sigma}^{\rm M}_{yy}$ overall decreases with contracting $a$, which in turn decreases the magnitude of ${\Delta}E_{\rm 8{\times}2-4{\times}1}$ as 28.5, 24.1, and 18.1 meV per 4${\times}$1 unit cell at $|{\Delta}a|/a_0$ = 0.3, 0.6, and 1\%, respectively. Thus, we can say that both the ES induced by electron doping and the MS induced by lattice contraction equally contribute to tune the structural phase stabilization in the In/Si(111) system.

\begin{figure}[ht]
\centering{ \includegraphics[width=7.7cm]{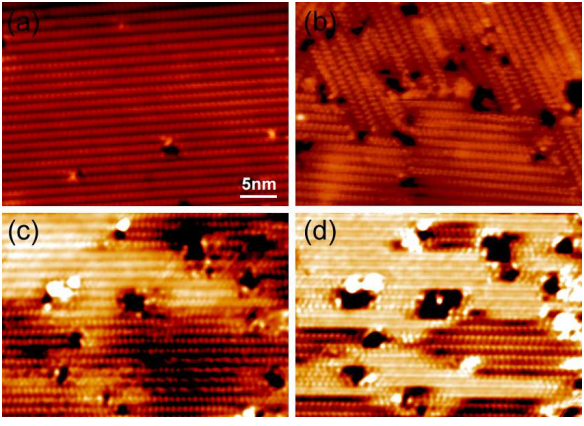} }
\caption{(Color on line)  STM images of the In atom wires on Si(111) substrates. (a) Defect-poor In wires on n-type Si (defect density of 0.0038 nm$^{-2}$), $V_{s}$ $=$ $-$1 V, $I_{t}$ $=$ 5 pA. (b) Defect-rich In wires on p-type Si (defect density of 0.0085 nm$^{-2}$), $V_{s}$ $=$ $-$1 V, $I_{t}$ $=$ 2 pA. (c),(d) The same area of defect-rich In wires on n-type Si (defect density of 0.0086 nm$^{-2}$), with $V_{s}$ $=$ $-$1 V and 1 V, respectively and the same $I_{t}$ of 2 pA. $V_{s}$ and $I_{t}$ denote the sample bias and tunneling current, respectively. All the STM images are of the same size and were acquired at 5 K.}
\end{figure}

In order to verify our theoretical prediction of the MS- and ES-driven tuning effects on the stabilities of competing 4${\times}$1 and 8${\times}$2 phases, we have performed STM experiments at 5 K~\cite{expt}.
We intentionally created vacancy defects in In wires to induce strain fields. It is observed that only the 8${\times}$2 phase exists at low defect density [Fig. 6(a)], while both the 4${\times}$1 and 8${\times}$2 phases coexist at high defect density [Fig. 6(c)]. The latter electronically phase-separated ground state can be attributed to large compressive strains~\cite{Zhang3} due to high defect density, consistent with the present theoretical prediction that the magnitudes of ${\Delta}{\sigma}^{\rm M}_{xx}$+${\Delta}{\sigma}^{\rm M}_{yy}$ and ${\Delta}E_{\rm 8{\times}2-4{\times}1}$ decrease with contracting $a$. On the other hand, the ES-driven tuning effect is demonstrated by adopting either $n$- or $p$-type substrate: i.e., for a certain defect density, only the 8${\times}$2 phase is present on $p$-type substrate (hole doping) [Fig. 6(b)], whereas both the 4${\times}$1 and 8${\times}$2 phases coexist on $n$-type substrate (electron doping) [Fig. 6(c)]. Alternatively, we utilize the surface charging effect at 5 K, where carrier relaxation between the surface layer and the bulk is substantially suppressed. As shown in Fig. 6(c) and 6(d), negative bias voltage (hole doping) tends to stabilize the 8${\times}$2 phase, while positive bias voltage (electron doping) favors the 4${\times}$1 phase. Therefore, it is demonstrated that increasing electron doping can favor the stabilization of the 4${\times}$1 structure.

\vspace{0.4cm}
\section{SUMMARY}
\vspace{0.4cm}

We have demonstrated the equivalent roles of ES and MS in tuning the relative stability of the 4${\times}$1 and 8${\times}$2 structures in the In/Si(111) surface. By means of hybrid DFT+vdW calculation, we found that electron doping into In wires for the 4${\times}$1 and 8${\times}$2 structures induces the ES with drastically different features, leading to a decrease in the surface energy difference between the two structures. We also found that applying a compressive lattice strain yields similar results for the surface-stress and surface-energy differences between the 4${\times}$1 and 8${\times}$2 structures. The equivalent control of phase stability by ES and MS has also been validated by low-temperature STM experiments. The present findings have important implications in understanding the underlying driving forces involved in various phase transitions of simple and complex systems alike, as well as in tailoring the physical properties of such systems.

\vspace{0.4cm}
\noindent {\bf ACKNOWLEDGEMENTS}
\vspace{0.4cm}

This work was supported in part by National Research Foundation of Korea (NRF) grant funded by the Korea Government (NRF-2011-0015754 and Grant No. 2014M2B2A9032247). C.Z. and Z.Z. acknowledge
support from NSFC (Grants Nos. 11434009, 11374279, 11461161009), NKBRPC  (Grant Nos. 2014CB921101, 2014CB921102), CAS  (Grant No. XDB01020000), and FRFCU (Grants No.  WK2340000011). J. Y. acknowledges
support from NBRPC (Grant No.2012CB921300). The calculations were performed by KISTI supercomputing center through the strategic support program (KSC-2014-C3-011) for the supercomputing application research.

\vspace{0.4cm}
\noindent {\bf APPENDIX: EVALUATION OF THE SURFACE MS DIFFERENCE BETWEEN THE 8${\times}$2 AND 4${\times}$1 STRUCTURES}
\vspace{0.4cm}

The surface energy (${\gamma}$) per unit area of the In/Si(111) surface system can be defined using a centrosymmetric slab geometry of which both sides consist of two equivalent surfaces~\cite{Stekolnikov}:

\begin{eqnarray}
{{\gamma} = \frac{1}{2A}[E_{slab}^{sym} - \mu_{Si}N_{Si} - \mu_{In}N_{In}],}
\end{eqnarray}

\noindent where $A$ is the surface area of the unit cell, ${\mu}_{Si}$ (${\mu}_{In}$) is the Si (In) chemical potential, $i.e.$ the energy per atom in bulk, $N_{Si}$ ($N_{In}$) is the number of Si (In) atoms in the unit cell, and $E_{slab}^{sym}$ is the total energy of the centrosymmetric slab. The factor of 1/2 is introduced to take into account the presence of two surfaces in the symmetric slab. Thus, the surface energies of the 8${\times}$2 (${\alpha}$) and 4${\times}$1 (${\beta}$) structures are given as:

\begin{eqnarray}
{{\gamma}_{\alpha} = \frac{1}{2A_{\alpha}}[E_{slab,\alpha}^{sym} - 4\mu_{Si}N_{Si} - 4\mu_{In}N_{In}]}\\
{{\gamma}_{\beta} = \frac{1}{2A_{\beta}}[E_{slab,\beta}^{sym} - \mu_{Si}N_{Si} - \mu_{In}N_{In}]}.
\end{eqnarray}

 The MS difference {$\Delta\sigma^{M}_{ij}$} between the 8${\times}$2 and 4${\times}$1 structures can be defined~\cite{Camm} as

\begin{eqnarray}
{\Delta\sigma_{ij}^{\rm M}} = {\sigma}_{ij,{\alpha}}^{\rm M} - {\sigma}_{ij,{\beta}}^{\rm M} = \frac{1}{A_{\alpha}} \frac{{\partial}(A_{\alpha}{\gamma}_{\alpha})}{{\partial}{\epsilon}_{ij}} - \frac{1}{A_{\beta}} \frac{{\partial}(A_{\beta}{\gamma}_{\beta})}{{\partial}{\epsilon}_{ij}}\\
= \frac{1}{2A_{\alpha}} \frac{{\partial}E_{slab,\alpha}^{sym}}{{\partial}{\epsilon}_{ij}} - \frac{1}{2A_{\beta}} \frac{{\partial}E_{slab,\beta}^{sym}}{{\partial}{\epsilon}_{ij}},
\end{eqnarray}
where $\epsilon_{ij}$ is the surface strain tensor ($i, j$ = $x, y$). The last equality holds since the stress of the bulk at the equilibrium lattice constant is zero. Assuming that the surface geometry obtained using the symmetric slab is the same as that obtained using the H-terminated slab, $\frac{1}{2}$($E_{slab,\alpha}^{sym}-\frac{A_{\alpha}}{A_{\beta}}E_{slab,\beta}^{sym}$) is equal to $E_{slab,\alpha}-\frac{A_{\alpha}}{A_{\beta}}E_{slab,\beta}$, where $E_{slab}$ is the total energy obtained using the H-terminated slab. Therefore, Eq. (7) can be expressed in terms of the first order change in the total energy of the H-terminated slab:

\begin{eqnarray}
{\Delta\sigma_{ij}^{\rm M}} = \frac{1}{A_{\alpha}} \frac{{\partial}E_{slab,\alpha}}{{\partial}{\epsilon}_{ij}} - \frac{1}{A_{\beta}} \frac{{\partial}E_{slab,\beta}}{{\partial}{\epsilon}_{ij}}
\end{eqnarray}

\noindent Corresponding authors: $^{\dagger}$chojh@hanyang.ac.kr, $^{*}$cgzeng@ustc.edu.cn



\begin{thebibliography}{99}
\bibitem{SCL} For electron doping, a space-charge layer is formed by charge transfer from bulk dopants. Since the present result of ${\Delta}E_{\rm 8{\times}2-4{\times}1}$ as a function of $n_e$ is well comparable with the STM and ARPES experiment~\cite{Morikawa} (see Fig. 4), it is likely that the formation energy of a space-charge layer minimally influences the energy difference ${\Delta}E_{\rm 8{\times}2-4{\times}1}$ between the 8${\times}$2 and 4${\times}$1 structures or their stress difference.
\bibitem{expt} The experiments were performed in ultrahigh-vacuum systems equipped with low temperature STM. Two types of Si(111) substrates were used in the experiments: As-doped n-type Si with room temperature resistivity of ${\sim}$ 0.002 ${\Omega}$${\cdot}$cm and boron-doped p-type Si with room temperature resistivity of ${\sim}$ 0.02 ${\Omega}$${\cdot}$cm. The substrates were cleaned following standard procedures by flashing at ${\sim}$ 1250${\celsius}$ and then cooling down slowly from 900${\celsius}$ to develop 7${\times}$7 reconstruction on the surfaces. The clean Si surfaces were confirmed by STM. Subsequently, about one monolayer of In was evaporated from a effusion cell onto the clean Si surfaces at ${\sim}$ 700 K followed by post-annealing at the same temperature for about half an hour to form In atom wires on the surface. Different effusion-cell temperatures were adopted to regulate the defect density in the In wires: lower (higher) effusion-cell temperature gives rise to higher (lower) defect density (see Ref. [12] for more details). After growth, the samples were transferred $in$ $situ$ for STM characterizations.
\end{thebibliography}
\end{document}